\documentclass[floatfix,superscriptaddress]{revtex4}
\usepackage{graphicx}
\usepackage{dcolumn}
\usepackage{amsmath}
\usepackage{epsfig}
\usepackage{multirow}

\graphicspath{{figures/}}

\newcommand{\BABARPubYear}{13}
\newcommand{\BABARPubNumber}{01}
\newcommand{\SLACPubNumber}{15451}

\newcommand{\BaBarType}       {CONF}

\input babarsym
\newcommand{\pvec}{{\bf p}}
\newcommand{\half}{\mbox{$\frac{1}{2}$}}
\newcommand{\bei}{\begin{itemize}}
\newcommand{\eei}{\end{itemize}}
\newcommand{\beq}{\begin{equation}}
\newcommand{\eeq}{\end{equation}}
\newcommand{\beqn}{\begin{eqnarray}}
\newcommand{\eeqn}{\end{eqnarray}}
\newcommand{\beqns}{\begin{eqnarray*}}
\newcommand{\eeqns}{\end{eqnarray*}}

\newcommand{\bkkkboth}{\ensuremath{\Bp \rightarrow \Kp\Km\Kp}\xspace}

\newcommand{\kkk}{\ensuremath{\Kp\Km\Kp}\xspace}

\newcommand{\de}{\ensuremath{\Delta E}\xspace}

\newcommand{\phiI}{\ensuremath{\phi(1020)}\xspace}

\def\Ampbar{\kern 0.18em\overline{\kern -0.18em {\cal A}}}

\newcommand{\optbar}[1]{\shortstack{{\tiny(\rule[.4ex]{.8em}{.2mm})}\\[-.7ex]$#1$}}
\def\AorAbar {\kern 0.18em\optbar{\kern -0.18em {\cal A}}{}\xspace}

\def\BzorBzbar    {\kern 0.18em\optbar{\kern -0.18em \Bz}{}\xspace}

\newcommand{\Acp}{\ensuremath{A_{\CP}}\xspace}

\def\splot{\ensuremath{_s{\cal P}lot}\xspace}

\newcommand{\gevcccc}{\ensuremath{{\mathrm{\,Ge\kern -0.1em V^2\!/}c^4}}\xspace}

\long\def\inst#1{\par\nobreak\kern 4pt\nobreak
    {\it #1}\par\vskip 10pt plus 3pt minus 3pt}

\begin{document}

\begin{flushleft}
SLAC-PUB-\SLACPubNumber \\
\babar-\BaBarType-\BABARPubYear/\BABARPubNumber \\[2mm]
\end{flushleft}


\title{
\large \bf
\boldmath
Study of the \Kp \Km invariant-mass dependence of \CP asymmetry in \bkkkboth decays
} 

%
\author{J.~P.~Lees}
\author{V.~Poireau}
\author{V.~Tisserand}
\affiliation{Laboratoire d'Annecy-le-Vieux de Physique des Particules (LAPP), Universit\'e de Savoie, CNRS/IN2P3,  F-74941 Annecy-Le-Vieux, France}
\author{E.~Grauges}
\affiliation{Universitat de Barcelona, Facultat de Fisica, Departament ECM, E-08028 Barcelona, Spain }
\author{A.~Palano$^{ab}$ }
\affiliation{INFN Sezione di Bari$^{a}$; Dipartimento di Fisica, Universit\`a di Bari$^{b}$, I-70126 Bari, Italy }
\author{G.~Eigen}
\author{B.~Stugu}
\affiliation{University of Bergen, Institute of Physics, N-5007 Bergen, Norway }
\author{D.~N.~Brown}
\author{L.~T.~Kerth}
\author{Yu.~G.~Kolomensky}
\author{M.~J.~Lee}
\author{G.~Lynch}
\affiliation{Lawrence Berkeley National Laboratory and University of California, Berkeley, California 94720, USA }
% ...
\collaboration{The \babar\ Collaboration}
\noaffiliation


\begin{abstract}
   \noindent
   As a followup to the latest ${\mbox{\slshape B\kern-0.1em{\smaller A}\kern-0.1em B\kern-0.1em{\smaller A\kern-0.2em R}}}$ amplitude analysis of the decay $B^+ \rightarrow K^+ K^- K^+$, we investigate the $K^+ K^-$ invariant-mass dependence of the $C\!P$ asymmetry and compare it to that obtained by the LHCb collaboration. The results are based on a data sample of approximately $470 \times 10^6 B\overline{B}$ decays, collected with the ${\mbox{\slshape B\kern-0.1em{\smaller A}\kern-0.1em B\kern-0.1em{\smaller A\kern-0.2em R}}}$
detector at the PEP-II asymmetric-energy $B$ factory at the SLAC National Accelerator Laboratory.  
\end{abstract}


\maketitle

A study of \CP violation in a Dalitz-plot analysis of \bkkkboth decays was performed by the \babar\ collaboration~\cite{Lees:2012kxa}. Based on this existing analysis, we exploit the \splot technique~\cite{Pivk:2004ty} to investigate the $K^+K^-$ invariant-mass dependence of the \CP asymmetry, $\Acp=\frac{\Gamma(\Bm) - \Gamma(\Bp)}{\Gamma(\Bm) + \Gamma(\Bp)}$.
The dependence of the \CP asymmetry on $K^+K^-$ invariant mass is
compared to a recent preliminary result from the LHCb collaboration~\cite{LHCB:2012-018}, where the direct \CP asymmetry in  $B^+\to  K^+K^-K^+$ over the entire phase space excluding charm decays was measured to be
\begin{equation}
\label{eq:Acp1}
\Acp(B^+\to K^+K^-K^+)=-0.046\pm 0.009 ({\rm stat.}) \pm 0.005({\rm syst.}) \pm 0.007 (J/\psi K^\pm).
\end{equation}
The first quoted uncertainty is statistical, the second is systematic, and the third is due to the uncertainty on
the measured value of the \CP asymmetry in $B‏\to\jpsi\Kpm$ decays (see below).
This result has a significance of 3.7$\sigma$ to be non-zero and is claimed to be the first evidence of \CP violation observed in inclusive charmless \B decays.  
The corresponding measurement from \babar\ is
\begin{equation}
\label{eq:Acp2}
\Acp(B^+\to K^+K^-K^+)=-0.017^{+0.019}_{-0.014} ({\rm stat.}) \pm 0.014({\rm syst.}),
\end{equation}
where no significant \CP violation is observed, although it is not inconsistent with the result from LHCb.  

The analysis method used to extract $\Acp$ is rather different between the experiments. \babar\ performs an amplitude analysis, based on a maximum-likelihood fit to the Dalitz plot
as well as the output of a neural network based on event shape variables
and the kinematic variables $\mes$ and $\de$~\cite{Lees:2012kxa}.  The energy-substituted mass is defined as
$\mes\equiv\sqrt{(s/2+{\mathbf {p}}_i\cdot{\mathbf{p}}_B)^2/E_i^2-p_B^2}$ 
and the energy difference $\de \equiv E_B^*-\half\sqrt{s}$, where 
$(E_B,\pvec_B)$ and $(E_i, \pvec_i)$ are the four-vectors of the \B 
candidate and the initial electron-positron system measured in the laboratory
frame, respectively.  The asterisk denotes the \epem CM frame, and $s$
is the invariant mass squared of the electron-positron system.  
Signal events peak at the \B mass ($\approx 5.279 \gevcc$) for \mes, 
and at zero for \de.
The inclusive $\Acp$ is calculated by separately integrating over the Dalitz plane
the efficiency-corrected charmless
isobar amplitudes for $B^+$ and $B^-$.  
The LHCb result is obtained by
fitting the $\Kp\Km\Kp$ and $\Km\Kp\Km$ invariant mass distributions, integrated over the Dalitz plot without any efficiency correction, and
calculating $\Acp^{RAW}=\frac{N^- - N^+}{N^- + N^+}$.
This raw asymmetry is corrected  by their observed $\jpsi \Kpm$ asymmetry of $-0.014\pm0.007$  to subtract residual charge asymmetries in production and detection.
This correction uses the world-average measured asymmetry of $0.001\pm0.007$ \cite{Beringer:1900zz} for $\Bpm \to \jpsi \Kpm$.
In the LHCb analysis, to remove contributions from the charm decays $\Bpm\to\Dzb(\Dz)h^{\pm}$ (where $h$ stands for $K$ or $\pi$) with $\Dzb(\Dz)\to h^+h^-$,
a $m_{\Kp\Km}$ veto was applied at $\pm 30\mevcc$ around the $\Dz$-mass value.
The inclusive \Acp extracted by LHCb is the integral over all the observed events in the $\kkk$ Dalitz plane. Unlike \babar, LHCb does not include a correction for varying efficiency across the phase space, but evaluates a systematic uncertainty of $0.15\%$ due to this effect.

\begin{figure}[htbp]
\includegraphics[width=7.0cm,keepaspectratio]{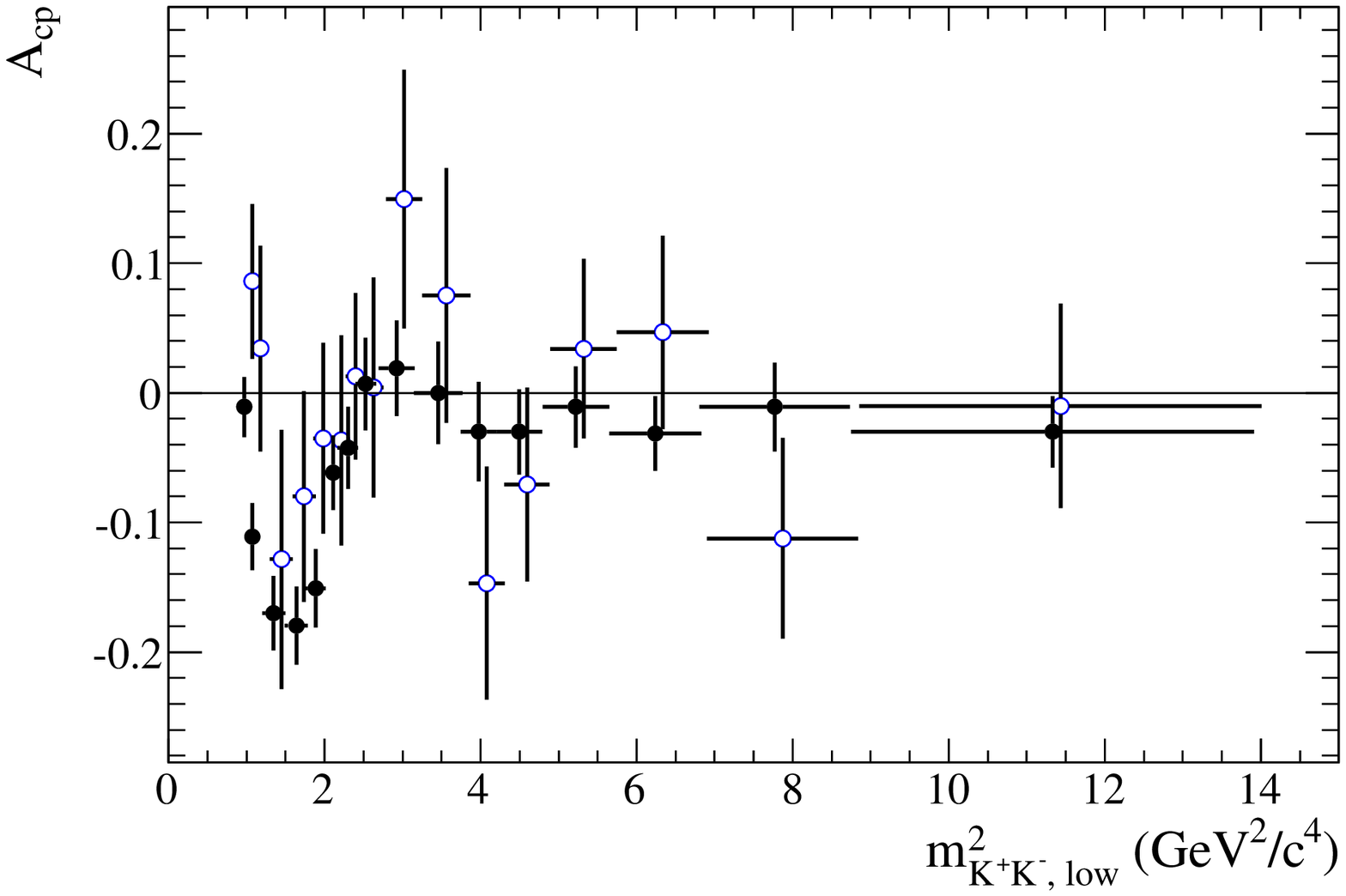}
\includegraphics[width=7.0cm,keepaspectratio]{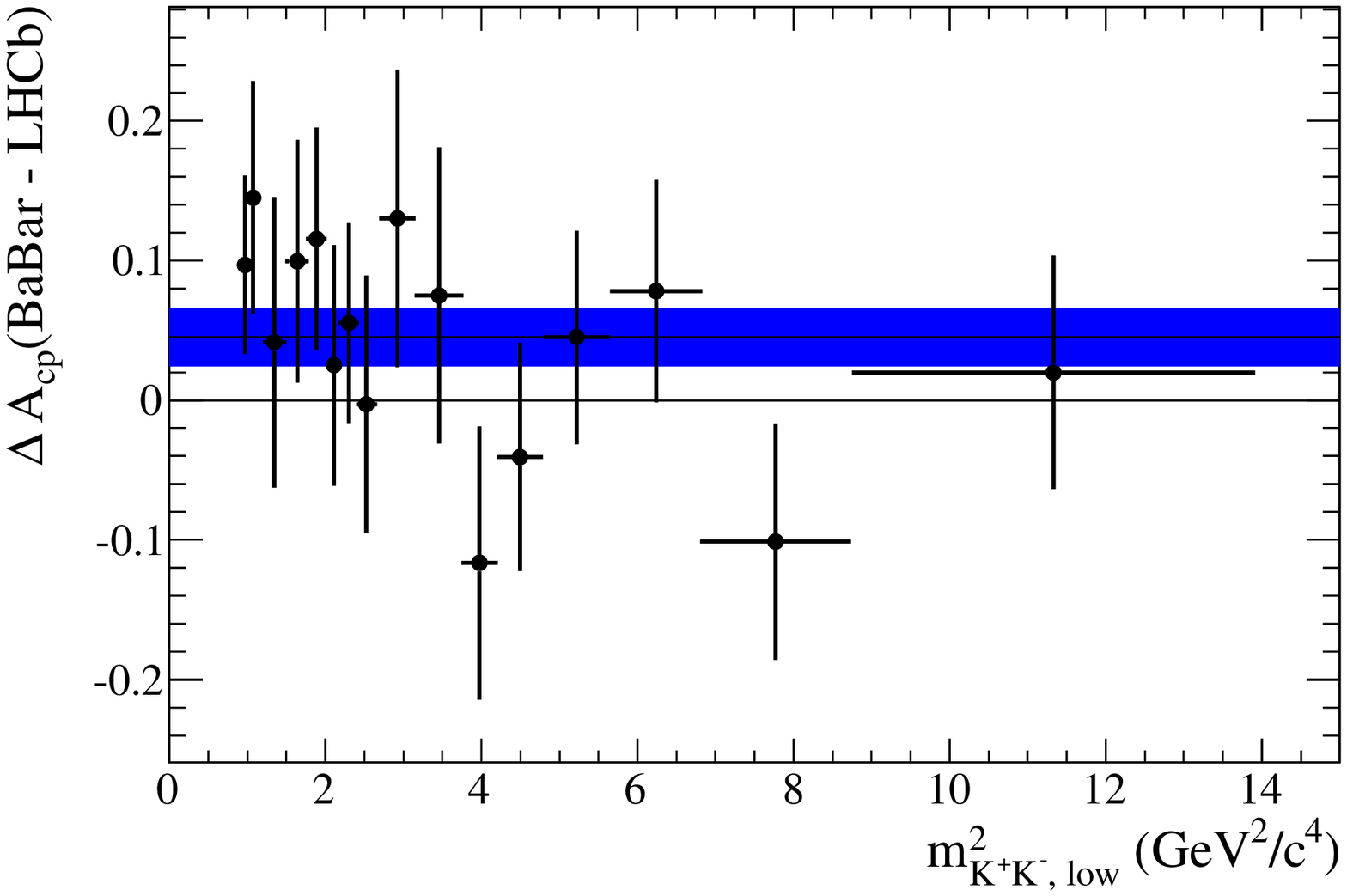}
\caption{\label{fig:lowKK}  Left:  $\Acp$ as a function of $m^2_{K^+K^-,{\rm low}}$ in $B^+\to K^+K^-K^+$ 
from LHCb (solid dots) and \babar\ (open dots).
The LHCb distribution is $\Acp^{RAW}$. The distribution from \babar\ is obtained by the \splot technique. For both experiments the error bars are statistical only. The systematic effects for \babar\ are estimated to be approximately $0.01$.
The \babar\ data points on the plot are shifted to the right by $0.1\gevcccc$ for clarity.
Right: The difference between the \babar\ and
LHCb asymmetries, $\Acp(\babar) - \Acp^{RAW}(LHCb)$.
Also shown is the average shift of $0.045\pm0.021$.
}
\end{figure}

LHCb also obtained the raw asymmetry as a function of the squared $K^+K^-$ invariant mass.  They observe a broad structure in the asymmetry at
$m_{\Kp\Km}^2 \approx 1.6 \gevcccc$. peaking at $\Acp \approx -0.2$.
The \babar\ publication did not directly include this study, although Fig. 8 in the \babar\ paper shows the $m_{\Kp\Km}$ distributions for $B^+$ and $B^-$ separately.
In this note, we have reproduced the binning and Dalitz plot cuts of the LHCb study in order to directly compare the mass dependence of \Acp between the two experiments.
The\babar\ \Acp distributions were produced with the \splot technique, using the \mes and \de variables, which are not correlated to each other or to the $K^+K^-$ invariant mass.
In Fig.\ref{fig:lowKK}, we show the extracted charge asymmetry as a function of the lower of the two $K^+K^-$ masses, $m_{\Kp\Km, {\rm low}}$.

Although the errors on the \babar\ data are approximately 2 times larger than those of LHCb, the pattern of the \CP asymmetry as a function of $m^2_{\Kp\Km, {\rm low}}$  agrees very well.  The $\chi^2$ between the data is $16.1$ for $16$ bins.  There does appear to be, however, a clear overall shift between
the measured LHCb and \babar\ asymmetries,
as shown in the right hand plot of Fig. \ref{fig:lowKK}.  The average difference between the binned $\Acp$ measurements is $0.045\pm0.021$ and appears to be flat across the spectrum. To obtain this average, we weighted the binned \Acp values by their respective errors.

The $K^+K^-$ invariant-mass spectrum in the region
$1.3-1.7 \gevcc$ includes contributions from at least the $f_0(1500)$, $f^\prime_2(1525)$, and $f_0(1710)$, as well as a broad non-resonant contribution~\cite{Lees:2012kxa}.  Considering the many varying strong phases involved, as well as the differing quark content of the different resonances, it is not surprising to see significant direct \CP violation in this region of phase space.

\begin{figure}[htbp]
\includegraphics[width=7.0cm,keepaspectratio]{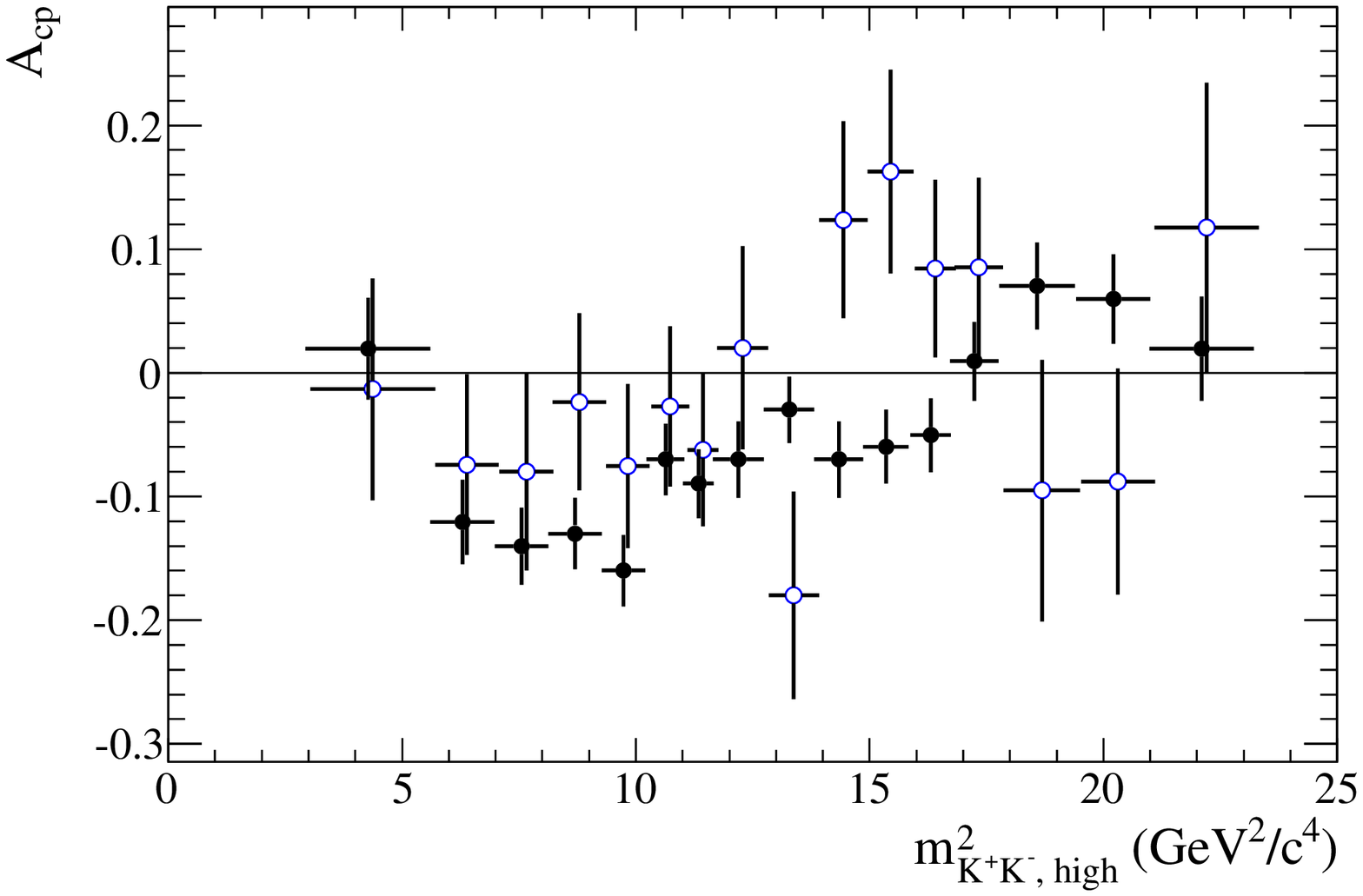}
\includegraphics[width=7.0cm,keepaspectratio]{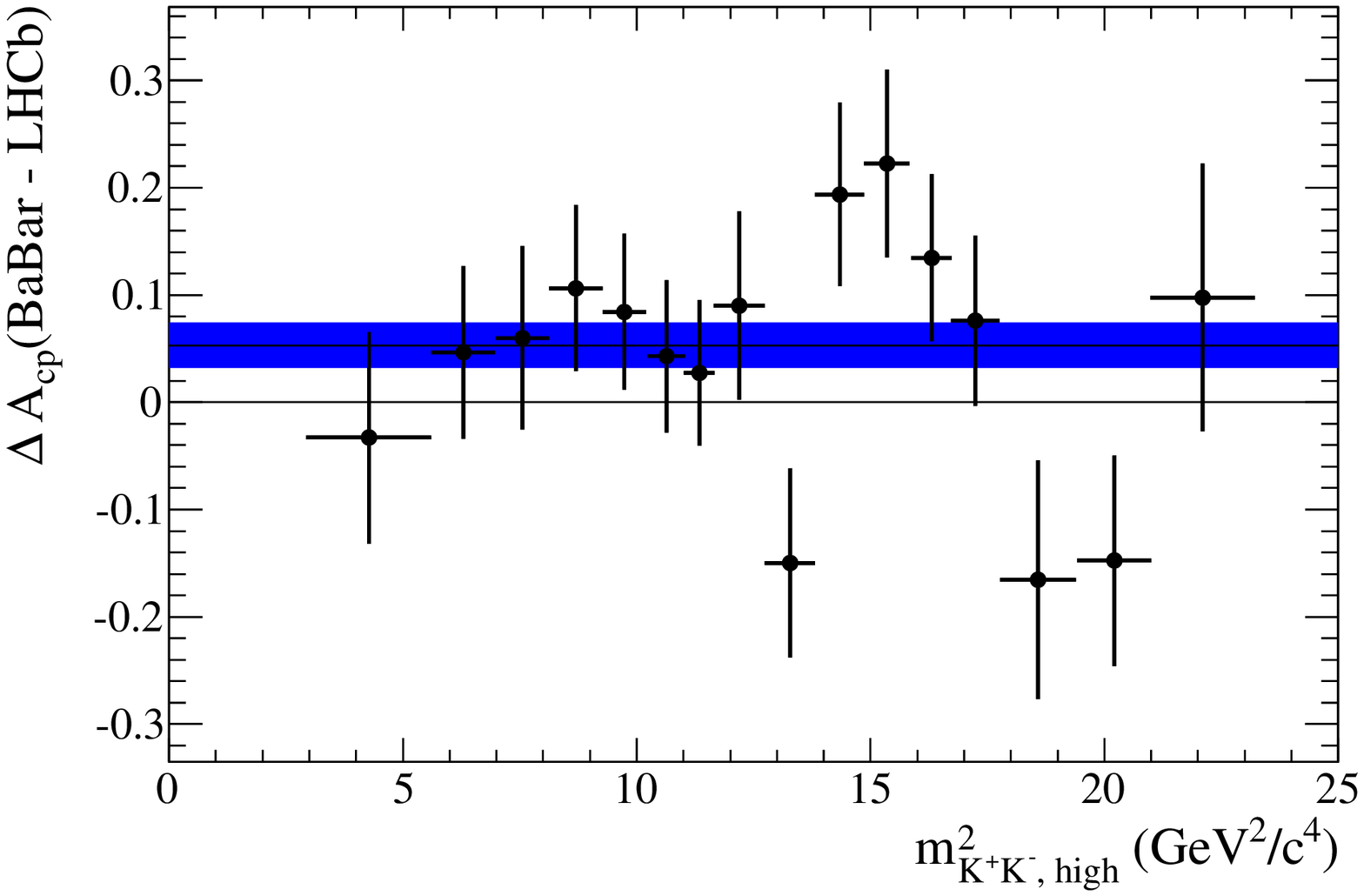}
\caption{\label{fig:hiKK}  Left:  $\Acp$ as a function of $m^2_{K^+K^-,{\rm high}}$ in $B^+\to K^+K^-K^+$ 
from LHCb (solid dots) and \babar\ (open dots).
The LHCb distribution is $\Acp^{RAW}$. The distribution from \babar\ is obtained by the \splot technique. For both experiments the error bars are statistical only.
The systematic effects for \babar\ are estimated to be approximately $0.01$.
The \babar\ data points on the plot are shifted to the right by $0.1\gevcccc$ for clarity.
Right: The difference between the \babar\ and
LHCb asymmetries, $\Acp(\babar) - \Acp^{RAW}(LHCb)$.
Also shown is the average shift of $0.053\pm0.021$.   
}
\end{figure}

For completeness, we also include similar plots the higher of the two $K^+K^-$ masses, $m_{\Kp\Km, {\rm high}}$, in Fig. \ref{fig:hiKK}. Here, the average shift is $0.053\pm0.021$.
The average shifts in asymmetry observed in $m_{\Kp\Km, {\rm low}}$ and $m_{\Kp\Km, {\rm high}}$ are similar but not identical. This behavior is expected due to the fact that we calculate the average of binned \Acp values weighted by the error and not by the number of signal events in each bin. The errors are influenced by the background distributions, which are different in the two variables. 

In summary, we performed a study of the $\Kp\Km$ invariant-mass dependence of the \CP asymmetry in \bkkkboth decays, based on a published \babar\ Dalitz-plot analysis~\cite{Lees:2012kxa}. The \babar\ data support the variation of the \CP asymmetry over the Dalitz plot seen by LHCb.
Nevertheless, a difference exists between the \CP asymmetries measured by \babar\ and LHCb. This difference appears to be consistent with being uniform across the phase space and is found to be $0.045\pm0.021$ between the \babar\ \Acp distribution as a function of $m_{\Kp\Km, {\rm low}}$ and that obtained by LHCb.
A compatible difference is observed in $m_{\Kp\Km, {\rm high}}$.
These values are consistent with the difference between the inclusive \Acp obtained by the two experiments.
The shift, while consistent with zero within $2$ standard deviation, explains the different conclusions between the two experiments concerning effects in specific regions of 
the phase space: the hint of direct \CP asymmetry in $\Bp\to\phiI\Kp$ 
that was seen by \babar\ but not confirmed by LHCb, and the fact that \babar\ finds a 
negative asymmetry with a smaller magnitude than LHCb around $m_{\Kp\Km}^2 \approx 1.6 \gevcccc$.
Further experimental investigation is needed to draw definitive conclusions on the source of \CP violation in \bkkkboth decays.

We are grateful for the excellent luminosity and machine conditions
provided by our \pep2\ colleagues, 
and for the substantial dedicated effort from
the computing organizations that support \babar.
The collaborating institutions wish to thank 
SLAC for its support and kind hospitality. 
This work is supported by
DOE
and NSF (USA),
NSERC (Canada),
CEA and
CNRS-IN2P3
(France),
BMBF and DFG
(Germany),
INFN (Italy),
FOM (The Netherlands),
NFR (Norway),
MES (Russia),
MINECO (Spain),
STFC (United Kingdom). 
Individuals have received support from the
Marie Curie EIF (European Union)
and the A.~P.~Sloan Foundation (USA).

\bibliography{ConfNoteKKK}

\begin{thebibliography}{4}
\expandafter\ifx\csname natexlab\endcsname\relax\def\natexlab#1{#1}\fi
\expandafter\ifx\csname bibnamefont\endcsname\relax
  \def\bibnamefont#1{#1}\fi
\expandafter\ifx\csname bibfnamefont\endcsname\relax
  \def\bibfnamefont#1{#1}\fi
\expandafter\ifx\csname citenamefont\endcsname\relax
  \def\citenamefont#1{#1}\fi
\expandafter\ifx\csname url\endcsname\relax
  \def\url#1{\texttt{#1}}\fi
\expandafter\ifx\csname urlprefix\endcsname\relax\def\urlprefix{URL }\fi
\providecommand{\bibinfo}[2]{#2}
\providecommand{\eprint}[2][]{\url{#2}}

\bibitem[{\citenamefont{{J.P. Lees} et~al.}(2012)}]{Lees:2012kxa}
\bibinfo{author}{\bibnamefont{{J.P. Lees}}} \bibnamefont{et~al.}
  (\bibinfo{collaboration}{BABAR Collaboration}), \bibinfo{journal}{Phys. Rev.}
  \textbf{\bibinfo{volume}{D85}}, \bibinfo{pages}{112010}
  (\bibinfo{year}{2012}).

\bibitem[{\citenamefont{Pivk and Le~Diberder}(2005)}]{Pivk:2004ty}
\bibinfo{author}{\bibfnamefont{M.}~\bibnamefont{Pivk}} \bibnamefont{and}
  \bibinfo{author}{\bibfnamefont{F.~R.} \bibnamefont{Le~Diberder}},
  \bibinfo{journal}{Nucl. Instrum. Methods Phys. Res.}
  \textbf{\bibinfo{volume}{A 555}}, \bibinfo{pages}{356}
  (\bibinfo{year}{2005}).

\bibitem[{\citenamefont{{The LHCb Collaboration}}(2012)}]{LHCB:2012-018}
\bibinfo{author}{\bibnamefont{{The LHCb Collaboration}}}
  (\bibinfo{year}{2012}), \bibinfo{note}{{Evidence for $C\!P$ violation in
  $B^{\pm}\to K^{\pm}\pi^+\pi^-$ and $B^{\pm}\to K^{\pm}K^+K^-$ decays,
  LHCb-CONF-2012-018, http://cds.cern.ch/record/1455471}}.

\bibitem[{\citenamefont{Beringer et~al.}(2012)}]{Beringer:1900zz}
\bibinfo{author}{\bibfnamefont{J.}~\bibnamefont{Beringer}} \bibnamefont{et~al.}
  (\bibinfo{collaboration}{Particle Data Group}), \bibinfo{journal}{Phys. Rev.}
  \textbf{\bibinfo{volume}{D86}}, \bibinfo{pages}{010001}
  (\bibinfo{year}{2012}).

\end{thebibliography}
\bibliographystyle{apsrev}
  
\end{document}